\documentstyle[twoside,fleqn,espcrc2]{article}

\newcommand{\AmS}{{\protect\the\textfont2
  A\kern-.1667em\lower.5ex\hbox{M}\kern-.125emS}}

\title{The Gluon Propagator in Momentum Space}

\author{C. Bernard\address{ Department of Physics, Washington University\\
St. Louis, MO 63130, USA}
        C. Parrinello\address{Physics Department, New York University\\
4 Washington Place, New York, NY 10003, USA\\
and Physics Department, Brookhaven National Laboratory\\
Upton, NY 11973, USA}\thanks{C.P. acknowledges financial support from C.N.R.}
and A. Soni\address{Physics Department, Brookhaven National Laboratory\\
Upton, NY 11973, USA}}

\begin{document}

\begin{abstract}
We consider quenched QCD on a $16^3 \times 40$ lattice at $\beta=6.0$ . We give
preliminary numerical results for the lattice gluon propagator evaluated both
in coordinate and momentum space. Our findings are compared with
earlier
results in the literature at zero momentum. In addition, by considering nonzero
momenta we attempt to extract the form of the propagator and compare it to
continuum predictions formulated by Gribov and others.
\end{abstract}

\maketitle

\section{INTRODUCTION}
The possibility of studying nonperturbatively on the lattice gauge-dependent
quantities provides in principle a unique tool to test QCD at the level of the
basic fields entering the continuum Lagrangian. From this point of view, the
gluon propagator in the quenched approximation is
perhaps the
simplest quantity. From its
study one expects to obtain among other things a better understanding of the
infrared behaviour of the theory
and of the mechanism of
gluon confinement.

The nonperturbative behaviour of the Euclidean gluon propagator has been
investigated  in the continuum by many authors with different methods and in
different gauges\cite{Gri,Cornwall,Stingl,Zwa,Nami}. In particular, a very
peculiar momentum dependence has been predicted  as a consequence of a
modification of the  standard path integral Faddeev-Popov formula in the Landau
gauge by the introduction of a  nonperturbatively correct gauge-fixing
procedure\cite{Gri,Zwa}. Such improved implementation of the Landau
gauge is expressed by the equations \begin{equation}
\partial \cdot A = 0 \qquad \qquad {\rm and} \qquad \qquad FP [A] > 0
\label{eq:gauge}
\end{equation}
where $FP [A]$ is the Faddeev-Popov operator in the Landau gauge,
which in
general is not positive definite. The positivity requirement in
(\ref{eq:gauge})
can be seen as a recipe to get rid (although not completely, see for
example\cite{over}) of Gribov copies\cite{Gri}. In the gauge (\ref{eq:gauge}),
the (transverse) gluon propagator in momentum space has been
argued to be of the form\cite{Gri,Zwa}
\begin{equation}
G_{Gribov} (k) \approx {k^2 \over k^4 + b^4}
\label{eq:prop}
\end{equation}
where $b$ is a dynamically generated mass scale. The form (\ref{eq:prop}) for
the propagator in momentum space implies that in the continuum
\begin{equation}
G_{Gribov} (\vec{k}=0,t) \approx e^{- {b \over \sqrt{2}} t} \ \left(
cos({b \over \sqrt{2}} t) - sin({b \over \sqrt{2}} t) \right)
\label{eq:propspace}
\end{equation}
 Remarkably, the same predictions were
also obtained
in the study of
Schwinger-Dyson equations\cite{Stingl}.

\section{THE LATTICE PROPAGATOR}

 The lattice gluon field can be defined as\cite{Mand}:
\begin{equation}
A_{\mu} (n) \equiv
{U_{\mu} (n) - U_{\mu}^{\dagger} (n)
\over 2 i a} - \frac{1}{3} tr \left(
{U_{\mu} (n) - U_{\mu}^{\dagger} (n)
\over 2 i a}  \right)
\label{eq:gluone}
\end{equation}
where $a$ is the lattice spacing. Thus the lattice gluon propagator in
$x-$space is the expectation value of:
\begin{equation}
G_{\mu \nu} (x, y) \equiv Tr \left(A_{\mu} (x) \ A_{\nu} (y) \right)
\label{eq:prolat}
\end{equation}

An important point is that on the lattice one can define and implement the
analogue of the gauge condition (\ref{eq:gauge}) and reobtain from analytical
arguments the predictions for the propagator mentioned in the above
section\cite{Zwa}.

In fact, given any link configuration $\{ U \}$, one can define a function of
the gauge transformations $g$ on $\{ U \}$
\begin{equation} F_U [g] \equiv - {1 \over V} \sum_{n, \mu} \ Re \ Tr \ \left(
U_{\mu}^{g} (n) + U_{\mu}^{g \dagger} (n- \hat{\mu}) \right), \label{eq:effecl}
\end{equation}
where $V$ is the lattice volume and $U^{g}$ indicates the gauge-transformed
link
$U_{\mu}^{g} (n) \equiv g (n) U_{\mu} (n) g^{\dagger} (n + \hat{\mu})$. An
iterative minimization of $F_U [g]$ obtained by performing suitable gauge
transformations generates a configuration $\{ U^{\bar{g}} \}$ that satisfies
the
lattice version of (\ref{eq:gauge}), defined in terms of a lattice
Faddeev-Popov
operator. Then it is natural to try and test numerically predictions like
(\ref{eq:prop}) and (\ref{eq:propspace}).

Numerical studies have been performed in the past years for the zero spatial
momentum Fourier transform of (\ref{eq:prolat}), namely $G(\vec{k}=0,t) \equiv
\sum_{i=1}^{3} \ G_{i i} (\vec{k}=0,t)$\cite{Mand,Gup,Soni}. These studies
reported some evidence of an effective gluon mass that increases with the time
separation. This feature,
which would be  unacceptable for the propagator of a
real physical particle
since it violates
the Kallen-Lehmann
representation, is in qualitative agreement with the continuum prediction
(\ref{eq:propspace}) and may be in principle acceptable for a confined
particle\cite{Stingl,Mand}.

Our work aims to test at a more quantitative level continuum predictions and
to
extend the above results through the study of the gluon
propagator at nonzero momenta.
By requiring consistency between zero and
nonzero momentum results,
one has a better chance to
determine the propagator's analytical form.

\section{NUMERICAL RESULTS}

\subsection{Technical Remarks}

It is perhaps worth
remarking that, unlike simulations involving quenched quark
propagators,
evaluations of purely gluonic correlation functions can take
full advantage of the translational symmetry of the theory in order to improve
statistics. On the other hand, such quantities turn out to be very sensitive
to the numerical accuracy of gauge fixing. Empirically, it turns out that when
the minimization of $F_U [g]$ has reached an accuracy $\approx .05 \%$ the
signal for the propagators is sufficiently stable against  additional gauge
fixing.

\subsection{Results}

\begin{figure}[htb]
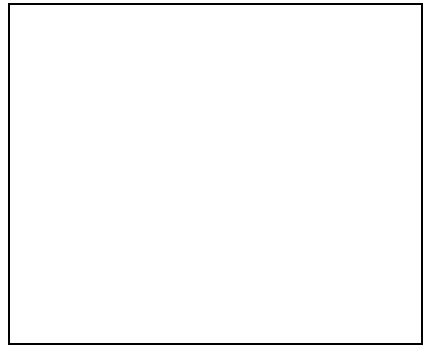
 \framebox[55mm]{\rule[-21mm]{0mm}{43mm}}
\caption{Effective gluon mass in lattice units}
\label{fig:effmass}
\end{figure}

We report results for a set of 25 configurations on a $16^3 \times 40$ lattice
at $\beta = 6.0$. As a first step we have evaluated $ G (\vec{k}=0,t)$; our
results confirm that the propagator exhibits a massive decay in time
with an effective mass $a*m(t) \equiv ln ({G (\vec{k}=0,t) \over G
(\vec{k}=0,t+a)})$ that increases with $t$. In Fig.\ref{fig:effmass}) we plot
$a*m(t)$ versus $t$ in lattice units with
jackknife
errors.
Assuming the value of the inverse lattice spacing $a^{-1} \approx 2.0 GeV$
at $\beta=6.0$, the effective gluon mass that we measure ranges approximately
between 200 and 800 MeV.

\begin{figure}[htb]
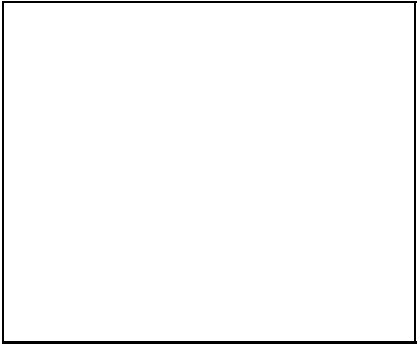

\framebox[55mm]{\rule[-21mm]{0mm}{43mm}} \caption{
 $G (\vec{k}=0,t)$ (error bars shown) and fit to the form
$G_{Gribov}$ (diamond points)}
\label{fig:fit1}
\end{figure}

We first attempt a 2-parameter least-squares fit
of our data to the continuum form
(\ref{eq:propspace})
without taking into account the correlations in the data.
The parameters are an overall normalization factor
and the mass scale $b$.
The results are given in Fig.
\ref{fig:fit1} and show a very good agreement between the data
and the fitting points.
For $a^{-1} = 2$ GeV one obtains $b = 225 \pm 5$ MeV,
where the quoted error
comes from
the covariance matrix of the fitting
parameters and does not include the systematic error on $a^{-1}$.

Good agreement is also
obtained by using the form commonly referred to as
particle $+$ ghost, that is $G (\vec{k}=0,t) \approx C_1 exp(-M_1 t) +
C_2 exp(-M_2 t)$,
where $C_2$ is constrained to be negative.

On the other hand, one cannot get good agreement
if one uses a conventional
4-parameter double exponential form,
that is if one constrains $C_2$ in the above
formula to take positive values.
Indeed, in this case the effective mass would always decrease
with $t$, in contrast to what is observed.

It is well known that the data points obtained from a Monte Carlo simulation
are in general statistically correlated; in the present case,
the correlated data
 are the
values of the propagator $G (\vec{k}=0,t)$ at different timeslices.

The standard way to take into account this effect when performing $\chi^2$ fits
is to use the definition of $\chi^2$ that involves the full covariance
matrix\cite{Tous}. Such a definition reduces to the standard one when the
covariance matrix is diagonal,
which
would happen if the data points were
uncorrelated.

By inspection of the covariance matrix for $G (\vec{k}=0,t)$ it turns out
that the off-diagonal matrix elements are typically of the same size as the
diagonal ones, i.e. our data points are
highly
correlated in $t$.
Consequently,
when we perform $\chi^2$ fits taking into account the full covariance matrix,
the fits are not well controlled because the correlation matrix is nearly
singular.
However, we
still
find
that $G_{Gribov}(\vec{k}=0,t) $ fits the data better than other
forms.
There is also qualitative
agreement between our results for $G(\vec{k}=0,t)$ and previous ones from other
groups\cite{Mand,Gup,Soni}.

\begin{figure}[htb]
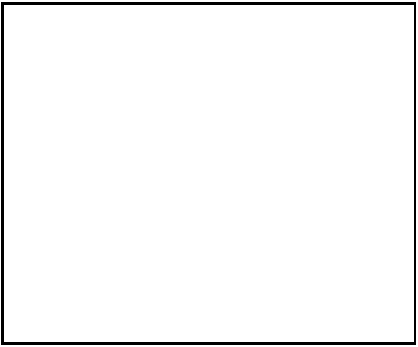
 \framebox[55mm]{\rule[-21mm]{0mm}{43mm}}
\caption{momentum space propagator vs. $|k|$ in GeV (assumes 1/a = 2.0 GeV)}
\label{fig:momprop} \end{figure}

In spite of the difficulties in the statistical analysis, our interpretation of
the results for $G(\vec{k}=0,t)$ receives a strong support from the analysis of
the momentum space propagator $G(k) \equiv \sum_{\mu=1}^4 G_{\mu \mu}(k)$. It
turns out that such a quantity is very well determined in a significant
interval of physical momenta, ranging from the lattice infrared cutoff $k_o =
{2 \pi \over N_t a}$ to $k \approx 3 k_o$  (see Fig. \ref{fig:momprop}). In
this range we fit the data to the continuum formula (\ref{eq:prop}) and, for a
comparison, to a standard massive propagator $G_{mass} (k) \approx {A \over
k^2 + m^2}$.

\begin{figure}[htb]
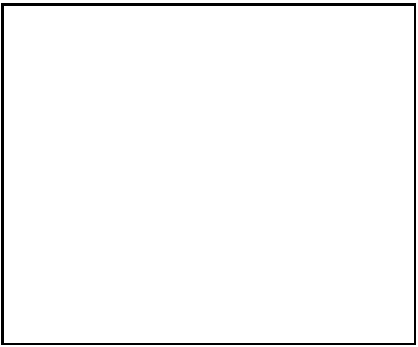

\framebox[55mm]{\rule[-21mm]{0mm}{43mm}}
\caption{
 $G (k)$ (error bars shown) and fit to the form $G_{Gribov}$ (diamond points)}
\label{fig:fit2}
\end{figure}

An
interesting
point is that the covariance matrix associated
with $G(k)$ turns out
to be much more
``diagonal" than the one for $G(\vec{k}=0,t)$; in other words,
the data points are much less correlated in momentum space than they are in
$t$.  As a consequence we have been able to obtain good fits for $G(k)$
with or without taking into account correlations.
We show in Fig. \ref{fig:fit2} a fit of $G(k)$ to the form $G_{Gribov} (k)$.
With
the full covariance
matrix, we get for the fit in Fig. \ref{fig:fit2} $\chi^2_{dgr} =
1.5$ and
$b = 322 \pm 8$ MeV,
assuming again $a^{-1} = 2$ GeV and neglecting systematic errors.

We compare this result to the best fit that one can obtain from the
standard massive propagator, for which we obtain $\chi^2_{dgr} =
2.9$.
On the other hand, the $b$ values that one obtains from the fits in coordinate
and momentum space differ
significantly, since they are respectively
$b = 225 \pm 5$ and $b = 322 \pm 8$ MeV.
Since an appreciable difference also occurs for the mass parameter $m$ when
we compare momentum and $x-$space fits to a simple massive propagator,
we think that
such discrepancies may be related to the different role that
finite-size effects play in the two calculations.
Further investigation of this issue is in progress.

\section{CONCLUSIONS}

We think that our results provide a significant (although preliminary) check of
the continuum predictions (\ref{eq:prop}) and (\ref{eq:propspace}). In
particular, the study of the propagator in momentum space appears very
promising since the data for such quantity turn out to be statistically
rather clean.

Recalling (\ref{eq:prop}) it is clear that a conclusive test requires the
study of the propagator at very low momenta, in order to observe the
suppression
predicted by (\ref{eq:prop}). Of course such a study calls for very big
lattices.

The work in progress aims to obtain first a better understanding of systematic
and statistical errors. After that, a study of the scaling properties of the
mass scale associated to the gluon propagator is in order and, in a later
stage,
the issue of the gauge dependence of the propagator will be addressed. \bigskip

C.P. wishes
to thank D. Zwanziger for many illuminating discussions.
C.B. was partially supported by the DOE under grant number
DE2FG02-91ER40628, and C.P. and
A.S. were partially supported under USDOE contract
number DE-AC02-76CH00016 .
The computing for this project was done at the National
Energy Research Supercomputer Center in part under the
``Grand Challenge'' program and at the San Diego Supercomputer Center.

\end{document}